\documentclass[twoside]{article}
\newdimen\mathindent
\AtEndOfClass{\mathindent\leftmargini}
\usepackage{espcrc2}

\usepackage{graphicx}
\usepackage[figuresright]{rotating}
\usepackage{caption2}

\font\af=msbm10 
\font\tenrm=cmr10 at 9truept
\font\eightpt=cmr8 at 8truept% roman at 8pt 
\font\rms=cmmi7 at 8truept%!!! should be 7pt
\font\sevenpt=cmr7 at 7truept 
\font\sevensy=cmsy7 at 7 truept % should be 7pt
\font\sevenrm=cmr9 \font\seveni=cmmi9 \font\sevensy=cmsy9
\font\fiverm=cmr8 \font\fivei=cmmi8 \font\fivesy=cmsy8
\font\fourrm=cmr7 \font\fouri=cmmi7 \font\foursy=cmsy7
\def\sevenpoint{\def\rm{\fam0\sevenrm}% switch back to 10-point type
\textfont0=\sevenrm \scriptfont0=\fiverm \scriptscriptfont0=\fourrm
\textfont1=\seveni  \scriptfont1=\fivei  \scriptscriptfont1=\fouri
\textfont2=\sevensy \scriptfont2=\fivesy \scriptscriptfont2=\foursy
\rm}

\def\ii{{\'{\i}}}
\def\th{{$\theta$\ }}
\def\tth{{$\theta$}}

\def\CE{{\cal E}}
\def\CO{{\cal O}}
\def\CM{{\cal M}}\def\CG{{\cal G}}
\def\CN{{\cal N}}
\def\cs{C_s(A)}
\def\be{\begin{eqnarray}}    
\def\ee{\end{eqnarray}}
\def\Dsl{\,\raise.15ex\hbox{/}\mkern-13.5mu D}
\def\CZ{{\cal Z}}
\def\Tr{{\rm Tr}}    
\def\tr{{\rm tr}}
\def\cp{{\af CP}$^{\rm N}$}

\def\cpa{{\af CP}$^{\rm 1}$} 

\def\IH{{\af H}}
\def\IZ{{\af Z}}
\def\Im{{\rm Im\ }}
\def\Re{{\rm Re\ }}

\newcommand{\AmS}{{\protect\the\textfont2
  A\kern-.1667em\lower.5ex\hbox{M}\kern-.125emS}}

\title{Vacuum energy and $\theta$--vacuum}

\author{M. Asorey \address{Departamento de F\ii sica Te\'orica, 
Facultad de Ciencias \\
Universidad de Zaragoza, 50009 Zaragoza, Spain}}
               
\begin{document}

\begin{abstract}
The highly non-trivial structure of the $\theta$--vacuum encodes many 
of the fundamental properties of gauge theories. In particular, 
the response of  the vacuum to the \tth--term perturbation
is sensitive to the existence of confinement, chiral symmetry 
breaking, etc. We analyze the dependence of  the vacuum energy 
density on \th around two special values, \th$=0$ and \th$=\pi$. The
existence or not of singular behaviors associated to spontaneous breaking 
of CP symmetry in these vacua  has been a controversial matter for years. 
We clarify this important problem by means of continuum non-perturbative 
techniques. The results show the absence of first order cusp singularities 
on the vacuum energy density at \th$=0$ and \th$=\pi$ for some gauge 
theories. This  smooth  dependence of the energy on \th  might have  
implications for long standing cosmological problems like the baryonic 
asymmetry and  the cosmological constant problem.
\vspace{1pc}
\end{abstract}

\maketitle

\section{Introduction}

In four-dimensional space-times non-abelian gauge theories exhibit
a highly non-trivial vacuum structure. A non-perturbative condensation
of classical configurations carrying special topological properties
(monopoles, vortices, instantons) provides a adequate medium for the
confinement of fundamental fermions. The special topological properties
of non-abelian groups and the dimensionality of space time make
also possible the appearance of a twist in the tunneling properties
of the vacuum. This twist of the vacuum can be explicitly induced
by the introduction in the action of a term  proportional to the
topological charge of the classical gauge fields. 
However, since thi term breaks CP invariance there are severe constraints 
in the value of its \tth--coefficient. In particular, the absence of a 
significant electric  dipolar momentum of the neutron requires that 
\th$ < 10^{-9}$ \cite{edn}.

There are two cosmological parameters with similar 
extremely small values: the baryon asymmetry
($\eta=N_B/N_\gamma\leq  10^{-10}$) which measures the ration of the 
number of baryons to the number of photons of the Universe, and the 
cosmological constant $\Lambda$ which although very small 
($\Lambda/(M_P^2)\leq 10^{-122}$ plays a fundamental role 
in the acceleration of the expansion of the Universe 
\cite{perl}. The unnatural smallness of these parameters has 
puzzled cosmologists and particle physicists for a long time
\cite{wein}.

In the last few years there has been an increasing interest on the 
solution of these longstanding cosmological problems.
Some proposals based on the 
structure of the \th vacuum  have been recently formulated \cite{yokoyama}.
In these scenarios the cosmological constant value  and the baryonic 
asymmetry are  associated to the tunneling between different 
classical vacua of a Yang-Mills like theory. The vacuum energy is assumed 
to vanish for some value of $\theta$ but the crucial assumption is that the
initial state of the Universe is instead of this $\theta$--vacua a state
localized around a classical vacua. The tunneling from this initial state to
$\theta$--vacuum is exponentially suppressed and might account for the 
tiny value of the present cosmological constant \cite{yokoyama} 
and the observed baryonic asymmetry \cite{muz}. According to this 
picture the  value of the cosmological constant will decrease 
with the time evolution of the Universe. Although the  approach is very
appealing it suffers from inconsistency. Gauge invariance implies
that all classical vacua are equivalent, thus, localization around one
specific vacua requires the breaking of gauge invariance. On the other
the existence of a finite mass gap requires that any physical state 
with very small energy must involve a fine-tuning of the coefficients 
of the series expansion of the state in stationary states.

From a different perspective a more conservative approach 
simply remarks the common  smallness of the three parameters
(QCD \th parameter, the cosmological constant and the baryonic asymmetry)
and suggests that there might exist an relation between them. 
One appealing feature of this scenario is that three fundamental
fine tuning problems are reduced to only one. Although one has  still 
to explain the reason for the existence of the common  fine tuning.

There are extra reasons to suspect that  \th is connected with the
other two cosmological parameters. In fact, the mere existence of 
a non-trivial value of \th$\neq 0$ has dramatic consequences for 
both cosmological problems. In all approaches to 
the solution of the baryonic asymmetry problem the requirement of 
violation of $CP$ symmetry appears as a fundamental prerequisite. 
Although there are other leading sources of CP violation the possibility 
of having a non-trivial \tth--vacuum is a complementary signal of the 
existence of the asymmetry.

The variation of the vacuum energy  with $\theta$ also opens the
possibility of a connection with the cosmological constant problem.
If the cosmological constant  vanishes for some value of 
\th (e.g. \tth$=0$) by some fundamental reason or 
theory (strings, supersymmetry, quantum gravity, etc),
an infinitesimal deviation of value of \th might induce a very tiny 
vacuum energy density enough to explain the small observed value of 
$\Lambda$. However, the constraint imposed on \th by the current bounds on 
the electric dipolar  momentum of the neutron are not enough to 
accommodate the smallness of the cosmological constant.
Indeed if we assume that the dependence of the vacuum energy on
\th is analytic around \th$=0$ 
the vacuum energy density $\CE_\theta$ of \tth--vacuum
is given by $\CE_\theta =\CE_0 + c\ \theta^2 + \CO(\theta^4)$, 
where the coefficient $c$ is proportional to the inverse of the 
mass of the lightest quark $\Lambda^6/m_{u}^2$.
%to the square of the mass of the $\eta'$ according to the 
%Witten-Veneziano formula\footnote{The coefficient c is in fact 
%proportional to the product of quark masses which is of the same 
%order of magnitude}. 
For \th$<10^{-9}$ we obtain a bound
$\CE=\CE_\theta - \CE_0< 10^{-22}\, {\rm GeV}^4$ which is much higher than
the cosmological bound  $\CE< 10^{ -47}\, {\rm GeV}^4$. Although the estimate
is very rough  the \tth--vacuum scenario requires to the  
bound on the value of \th to be more stringent \th$< 10^{-21}$ than 
that provided by the current measures of the  electric dipolar structure 
of the neutron. The estimate is based on the assumption that 
the vacuum energy density $\CE_\theta$ has a smooth dependence on \th\ 
%and the Witten-Veneziano formula holds. 
which   generically implies a lower variation of the vacuum energy density.

The smoothness of $\CE_\theta$ is claimed to hold  at first order 
around $\theta=0$ by the
Vafa-Witten theorem  \cite{vw} on the absence of spontaneous
symmetry breaking of CP symmetry in QCD. However,
some doubts  have been recently raised about the validity of the 
Vafa-Witten proof  of the theorem \cite{ag,ch,jw,Ji}.
In fact, $\CE_\theta$ is not always smooth at   
$\theta=\pi$: there are cases where CP is spontaneously broken at that   
point. In principle, the same pathology could appear also at   
$\theta=0$. 

If we assume that $\CE_\theta$ has a smooth behavior at 
$\theta=0$ and $\theta=\pi$ a very simple argument shows that the 
 expectation value of the topological density  
$$\CE'_\theta= -{\dot\imath\over 16 \pi^2}
\Big\langle F_{\mu\nu} \tilde{F}^{\mu\nu} 
\Big\rangle_{\theta}$$
vanishes at $\theta=0$ and $\theta=\pi$, i.e. $\CE'_{\theta=0}
=\CE'_{\theta=0}=0$.  
The argument is based on the fact that
the two classically CP symmetric vacua $\theta=0$ and $\theta=\pi$ 
are extremal points
of the vacuum energy density. In the case $\theta=0$ this property
follows from 
reflection CP symmetry $\CE_\theta=\CE_{ -\theta}$ which implies that
 $\CE_\theta$ has a local extrema (maximum or minimum) at
\th$=0$. In the case \tth--$\pi$ the same property
follows from Bragg symmetry 
$\CE_{\pi+\theta}=
\CE_{\pi-\theta}$, which is is a consequence of 
 CP symmetry  and  $\theta$-periodicity $\CE_\theta=\CE_{\theta+2\pi}$.
Thus, if  $\CE_\theta$ is smooth its first derivative 
and the vacuum expectation value  of the topological density 
vanish at $\theta=0$ and $\theta=\pi$. The vanishing of this  
order parameter of CP symmetry is natural because the
smoothness $\CE_\theta$ is equivalent to the absence of first order
phase transition and therefore compatible with the absence of 
spontaneous breaking.

Thus, the search for the smoothness of $\CE_\theta$ around
the points $\theta=0$ and $\theta=\pi$ becomes in fact
a search of spontaneous breaking of CP symmetry.
In this paper we further clarify the presence or not
of cusp singularities in $\CE_\theta$ around the 0--vacuum and
$\pi$--vacuum. Among the four possible scenarios with possible
cusps in any of the CP symmetric vacua (Fig. 1) we
will find that  QCD  realizes the  maximally smooth behavior
without cups (Fig. 1(d)).
\\

 \includegraphics[scale=.5]{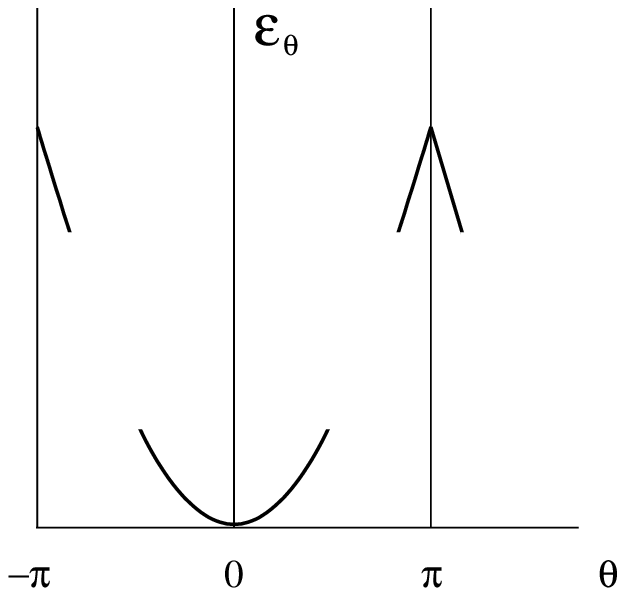}
 \includegraphics[scale=.5]{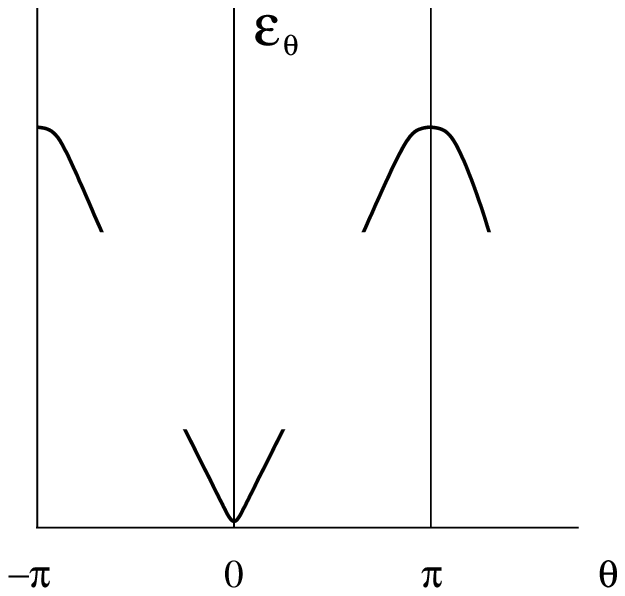} 
\vskip -4pt
 $\phantom{}$  \hskip1.4cm \hbox{ (a)} \hskip2.94cm \hbox{ 
(b)}
\vskip 8pt
 \includegraphics[scale=.5]{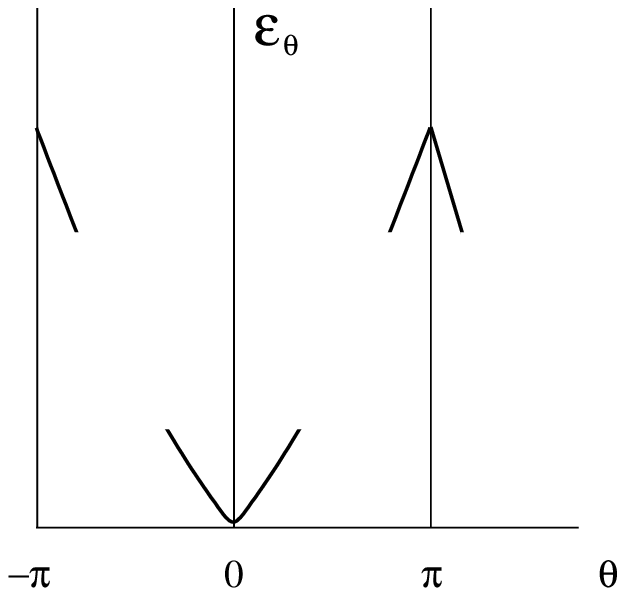} %$\, $
 \includegraphics[scale=.5]{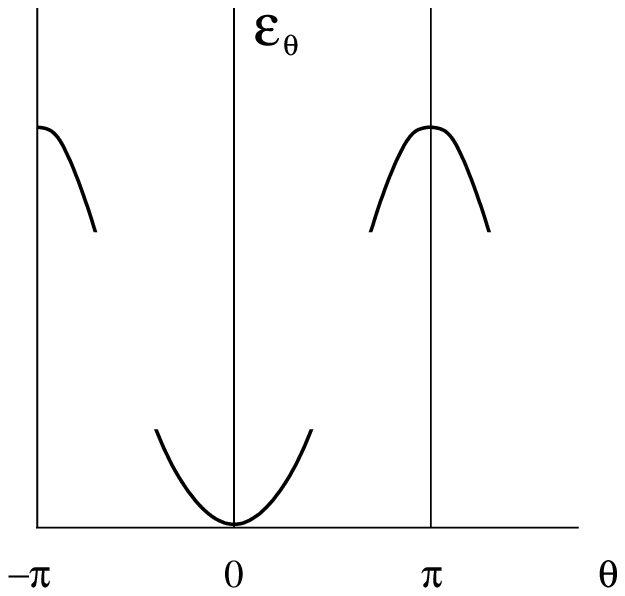}
\vskip -4pt
$\phantom{}$  \hskip1.5cm \hbox{(c)} \hskip2.95cm \hbox{ 
(d)}

\vskip 8pt
\baselineskip 0mm{\tenrm{ Figure 1.} 
\sevenpoint
Four different scenarios for possible behaviors of the vacuum energy
density  ${\CE_\theta}$. In the first case (a) the cusp at $\theta=\pi$ 
signals the spontaneous breaking of CP  symmetry in $\pi$--vacuum. 
In case (b)  CP symmetry is broken at ${\theta=0}$ but the $\pi$--vacuum is CP 
invariant. In the third case (c) there are two cusps at $\theta=\pi$ 
and $\theta=0$ meaning that  none of those vacua is CP invariant. Finally, 
in case (d)  $\CE_\theta$ is smooth and CP symmetry is
preserved for both values of \tth.}

\section{Vafa-Witten theorem}

\baselineskip 12pt
The Vafa-Witten theorem  is one of the very few    
non-perturbative analytic results of QCD. It provides a strong
constraint on the behavior of the vacuum energy density
$\CE_\theta$ for small values of \tth--parameter.
A very simple and elegant argument pointed by Vafa and Witten    
\cite{vw} shows why  spontaneous  violation    
of parity symmetry is not present in QCD
and  in any gauge theory with Dirac fermions in 3+1 space-time 
dimensions. This result implies, in particular, that
the expectation value of the topological density 
vanishes in the $\theta=0$ vacuum. The argument exploits 
the positivity of the Yang-Mills Euclidean measure \cite{amm,aff} and    
fermionic determinants \cite{vw} which implies the existence of a real    
effective action 
$$ S_{\rm eff}(g)= {1\over 2g^2}
\int\!  F_{\mu\nu} F^{\mu\nu} - \log\det  (\Dsl+m)$$ 
and an upper bound for partition function of the \tth--vacuum 
\begin{equation} 
\CZ_\theta =       
\int\! { \delta A}\,  {\rm e}^{- S_{\rm eff}(g)+ 
{\imath\theta\over 16 \pi^2} \int\!  F_{\mu\nu} \tilde{F}^{\mu\nu}}\!
< \CZ_0        
\label{ptf}    
\end{equation}    
in terms of the $0$--vacuum partition function. Notice that the 
partition function $\CZ_\theta$ is always real because
CP symmetry  transforms $F \tilde{F}$ into $ -F \tilde{F}$
and preserves the effective gauge action $ S_{\rm eff}(g)$. 
The inequality (\ref{ptf}) implies that the free energy density    
$\CE_\theta=E_\theta/V$ of the $\theta$--vacuum is bounded below by that  
of the $0$--vacuum, i.e. $\CE_0\leq \CE_\theta$; from this property 
Vafa-Witten argued implicitly assuming that
$\CE_\theta$ is smooth at \th$=0$  that $\CE'_\theta=\big<F
\tilde{F}\big>_0=0$. A criticism recently raised to the 
Vafa-Witten proof objects that the very existence of a first order 
phase transition might imply that the vacuum energy density or 
the free energy is not well defined \cite{ag} and,
thus, the Vafa-Witten argument will be invalid.
However, in this case there is no doubt about the existence of 
a well defined free energy for any real value of the $\theta$ parameter. 
This is guaranteed by unitarity of the theory, which is translated into    
Osterwalder-Schrader (OS) positivity of the Euclidean functional    
measure of the theory with real \tth--terms \cite{es,am}. 
This special property of Yang-Mills theory excludes the possibility    
that pathological scenarios where the free energy is ill-defined    
\cite{ag} might occur in the theory.
However, this fact does not  mean that  the theorem is already proven.    
One needs to exclude the existence of a cusp in the energy density at    
$\theta=0$. The appearance of the cusp would signal the existence of a    
first order phase transition and spontaneous breaking of parity at    
$\theta=0$ without violation of the Vafa-Witten inequality for free    
energies. The only extra property one needs to prove is smoothness of    
$\CE_\theta$ at $\theta=0$, which would make the    
existence of such a cusp  not possible.  In other terms, $\theta=0$ is    
always an absolute minimum of the free energy density but only if    
$\CE_\theta$ is smooth we will have $\CE'_0=0$. Only in that case one    
can make sure that the vacuum expectation value of the topological    
density vanishes and the Vafa-Witten theorem holds  for    
this particular order parameter. The smoothness of $\CE_\theta$ at 
$\theta=0$ has been first proved in Ref.\cite{agas}. Let us summarize 
the essential arguments of the proof.   
      
\section{Analytic Continuation 
         and Lee-Yang's zeros}
    
The proof  is based on the analysis of
possible existence of 
Lee-Yang singularities. Non-analyticities in the \th dependence
can be easily traced from the lack of an analytic continuation into    
the complex $\theta$ plane. Indeed, for complex values of 
\th  OS positivity (i.e. unitarity) is not preserved and there 
is no guarantee that the free energy is well defined, 
in which  case it does not make sense to speak about smoothness. 
Essentially, there are two    
possible ways in which the $\theta$ theory might give rise to    
pathological non-analyticities in the complex sector. One    
possibility is that the analytic continuation of the partition function
for complex values of \th\ is not defined at all, 
i.e.  the partition function itself becomes divergent.  
The other possibility is that the    
partition function exhibits a sequence of zeros converging to    
$\theta=0$ in the infinite volume limit. This is the scenario    
advocated by Lee and Yang \cite{ly} which signals in many cases    
the existence of a phase transition. 
   
For complex values of \th 
the partition function can be split 
into a sum of functional integrals over    
the different topological sectors    
\be    
\CZ_\theta (g)=\sum_{q=-\infty}^\infty    
{\rm e}^{-q\Im\theta +\dot \imath q\Re\theta}    
\int\mkern-18mu{\lower3ex\hbox{${}_{c_2(A)=q}$}}    
\mkern-14mu{ \delta A}\  {\rm e}^{-S_{\rm eff}(g,q)}        
\nonumber
\ee    
\noindent   
In a strict sense, the functional integral is UV divergent but it    
can be regularized in such a way that its positivity properties are    
preserved \cite{aff}. We can consider such a geometrical    
regularization whenever it is required. On the other hand, in the    
infinite volume limit $\CZ_\theta (g)$ vanishes unless the vacuum    
energy is renormalized to 0. But it is just the dependence on    
$\theta$ of the vacuum energy what we want to analyze. Therefore,    
we will consider throughout this note a compact space-time with large    
but finite volume $VT<\infty$.    

 On the other hand the \tth--term of the action is
bounded above by the  Yang-Mills action $S_{\rm YM}(q)$     
in every $q$-topological sector by the BPS bound 
$S_{\rm YM}(g,q)\geq 4 \pi^2|q|/g^2$.   
In the regularized theory, the coefficient of this linear bound increases   
with the UV regulating scale. In general, from positivity of fermionic    
determinants, Yang-Mills measure and the BPS bounds we have the    
inequality    
\be  
\left|\CZ_\theta(g) \right|  \displaystyle &\leq&    
\sum_{q=-\infty}^\infty     
{\rm e}^{|q\Im\theta| } \int\mkern-10mu{\lower2ex\hbox{${}_{c_2(A)=q}$}}    
\mkern-9mu{ \delta A}\  {\rm e}^{-S_{\rm eff}(g,q)}
\crcr
&\leq& \displaystyle    
\sum_{q=-\infty}^\infty    
\int\mkern-10mu{\lower2ex\hbox{${}_{c_2(A)=q}$}}    
\mkern-9mu{ \delta A}\  {\rm e}^{-S_{\rm eff}(\tilde{g},q)}=    
\CZ_0 (\tilde{g})\cr
\nonumber%{arth}    
\ee    
\noindent   
where $1/\tilde{g}^2=1/g^2- |\Im\theta|/4\pi^2$. This shift of the    
gauge coupling constant can be considered as a change of the    
renormalization point.  This shows that the partition function of    
the original theory with a complex \tth--term is bounded by the    
partition function of a similar theory with $\theta=0$ but with a    
different coupling constant. Since the theory at $\theta=0$ is    
unitary and renormalizable, its partition function $\CZ_0$ is    
finite and  from the above inequality it follows that    
$\left|\CZ_\theta\right|<\infty$ is also finite for small values    
of $\Im\theta$.  If in addition the theory is asymptotically free,    
then the renormalization of $g^2$ in the UV fixed point can absorb    
any value of $\Im\theta$, which implies an infinite radius of convergence    
of the sum over topological sectors in the complex $\theta$ plane. 
To make the argument more precise 
one should consider the regularized theory. The bare coupling constant 
$g$ has to be fine tuned    
according to the renormalization group to yield the appropriate continuum    
limit. But because of asymptotic freedom this coupling goes to    
zero as the UV regulator is removed and the shift from $g$ to    
$\tilde{g}$ induced by $\Im \theta$ simply implies a change in the    
effective scale of the continuum theory.
   
The only remaining possible source of non-analyticity is the presence
of Lee-Yang singularities, i.e.  zeros in the partition function 
$\CZ_\theta$ which could prevent the existence of a unique limit of 
the free energy $\log \CZ_\theta$ or its derivatives at $\theta=0$. 
 Uniform convergence of  the sum over topological sectors 
implies continuity of $\CZ_\theta$ on the complex.    
Now, since $\CZ_\theta > 0$ for  finite space-time volumes and real 
values of $\theta$, from  $2 \pi   
$-periodicity of $\CZ_\theta$  and  compactness of $[0,2\pi]$   
follows the existence of an open  strip covering the real \tth--axis    
where $\CZ_\theta\neq 0$. This implies that for finite volumes there    
is no phase transition for any real  value of \th. However, in the    
infinite volume limit $\CZ\searrow 0$ as $V \nearrow \infty$   
and the argument does not provide  information  about the critical 
behavior of the theory. 

Now, the positivity argument of Vafa and Witten 
can be used to show the absence of Lee-Yang zeros for  pure 
imaginary values of $\theta$ in vector-like gauge theories. Indeed, if 
\th\ is purely imaginary $\theta=\dot\imath\vartheta$, the contribution
of every topological sector is positive ($\Re \theta=0$) and the 
full partition function is real and strictly positive 
($\CZ_{\dot\imath\vartheta}>0$), 
i.e. $\CZ_{\dot\imath\vartheta}$ has no Lee-Yang zeros for any value 
of $\vartheta$. This property is essential for the proof  parity 
symmetry cannot be spontaneously broken.

The partition function $\CZ_\theta$  can be split  into two
terms   $\CZ_\theta=\CZ^+_\theta+\CZ^-_\theta$, encoding 
the contributions of parity even ($\CZ^+_\theta$) and parity odd 
($\CZ^-_\theta$) states. In the physical \tth--sector ($\Im \theta =0$), 
 $\CZ^{\pm}_\theta$ are defined  by $\CZ^{\pm}_\theta={1\over 2}
\CZ_\theta\pm {1\over 2}\CZ^P_\theta$, where  $\CZ^P_\theta$ 
is given by $\Tr \, P\, {\rm e}^{-T H_{\theta+2 \pi}}$ in terms of the  
Hamiltonian $H_{\theta+2 \pi}$ and  parity  operator $P$.
For general values of \th,   $\CZ^P_\theta$ is given by 
the standard functional integral  $\CZ_\theta$ but  
with the $\theta$ parameter  shifted to $\theta+2 \pi$ and  parity odd 
boundary conditions $A(-{{{\hbox{{\rms {T}}\eightpt{/2}}}}})=
[A^P({\hbox{{\rms {T}}\eightpt{/2}}})]^\phi$, where
$\phi$ is any gauge transformation and $A^P({\hbox{{\rms {T}}
\eightpt{/2}}})$ is the parity transform  of $A({\hbox{{\rms {T}}
\eightpt{/2}}})$, i.e. $A^P({\hbox{{\rms {T}}\eightpt{/2}}})=( 
A^P_0({\hbox{{\rms {T}}\eightpt{/2}}}),{A}_i^P({\hbox{{\rms {T}}
\eightpt{/2}}}))$ with $ A^P_0({\bf \vec{x}},
{\hbox{{\rms {T}}\eightpt{/2}}})=
A_0(-{\bf\vec{x}},{\hbox{{\rms {T}}\eightpt{/2}}}),
{A}_i^P({\bf \vec{x}},{\hbox{{\rms {T}}\eightpt{/2}}})= -{A}_i(-{\bf 
\vec{x}},{\hbox{{\rms {T}}\eightpt{/2}}})$ \cite{vw}.
The  functional integral approach provides an explicit 
analytic continuation of $\CZ^P_\theta$, $\CZ^+_\theta$ and $
\CZ^-_\theta$ to the complex  domain $|\Im \theta|< \theta_c$. 
Since  $\CZ_\theta>0$  for any real or imaginary value of \th\
in finite volumes, by continuity there exists an open domain 
 in the \tth--plane covering the real and imaginary axis which is
free of  Lee-Yang zeros for $\CZ_\theta$ and $\CZ^\pm_\theta$.
In that domain   $\CZ^\pm_\theta$ can be rewritten as
$\CZ^\pm_\theta={\rm e}^{-\xi_\pm(\theta)}$ in terms of two
analytic functions $\xi_+(\theta)$ and $\xi_-(\theta)$.

If there were a first order phase transition at $\theta=0$ 
with spontaneous parity symmetry breaking 
the expectation value of the CP breaking order parameter
\be
&-&\displaystyle \!\!\!\!\!  {\dot\imath\over 16 \pi^2} \Big\langle 
 F_{\mu\nu} \tilde{F}^{\mu\nu} \Big\rangle_{\theta=0}^-=
\lim_{VT\to \infty}{\xi'_-(\theta)\over VT}\Big|_{\theta=0}
\phantom{\Bigg[}\cr\cr
&=&\displaystyle \!\!\!\!
{\dot\imath\over 16 \pi^2}
\Big\langle F_{\mu\nu} \tilde{F}^{\mu\nu} \Big\rangle_{\theta=0}^+\!\!
=-\!\lim_{VT\to \infty} \!\!
{\xi'_+(\theta)\over VT}\Big|_{\theta=0}\!\!\!
\neq 0\phantom{\Bigg[}\cr
\nonumber%{unb}
\ee
would  not vanish.
Since  $\CZ_\theta(g)$ is real for imaginary values of $\theta=\dot\imath 
\vartheta$ the existence of a cusp associated to level crossing 
would imply that,
$ \xi_-(\dot\imath \vartheta)^\ast=\xi_+(\dot\imath \vartheta)$ and 
$$
 \CZ_{\dot\imath \vartheta}= 2 {\rm e}^{-\Re\xi_+(\dot\imath \vartheta)} 
\cos\, \Im \xi_+(\dot\imath \vartheta).
$$
Now, the CP symmetry breaking condition  requires that $\Im
\xi_+(\dot\imath \vartheta)$ be unbounded for imaginary values of
$\theta=\dot\imath \vartheta$ close to $\vartheta=0$ in the infinite
volume limit. This fact implies that if the system undergoes a first
order phase transition with parity symmetry breaking an increasing
number of Lee-Yang zeros must arise in the partition function $
\CZ_\theta$ for imaginary values of $\theta$ and large enough
volumes. The localization of zeros in the unit circle $|w|=1$ of the
complex plane $w={\rm e}^{i\theta}$ is reminiscent of the Lee-Yang
theorem for spin systems \cite{ly}.  However this cannot occur in
vector-like gauge theories where we have shown that for any finite
volume the partition function $ \CZ_\theta$ has no zeros on the
imaginary line $\Re \theta=0$.  Therefore, the system cannot undergo a
first order phase transition at $\theta=0$ with parity symmetry
breaking. In other terms, there is not a first order cusp in the
vacuum energy density $\CE_\theta$ at $\theta=0$, i.e. $\CE'_0=0$ and
the Vafa-Witten theorem holds.

The existence of any odd-order phase transition can be discarded for
the same reason because the function $\Im\xi_+(\dot\imath\vartheta)$,
which is odd in $\vartheta$, would be unbounded and $ \CZ_\theta$ would
have Lee-Yang zeros for some values of $\theta=\dot\imath\vartheta
$. However, even-order phase transitions cannot be excluded by these
arguments because such transitions will not require spontaneous parity
symmetry breaking. Recent numerical estimates provide some evidence on the
absence of second order transitions for a large family of gauge 
groups \cite{dv}.

In summary, there is now a  complete proof of the first order 
smoothness of the vacuum energy density at $\theta=0$:
the missing link in the Vafa-Witten argument about    
the vanishing of the topological charge order parameter. 

\section { $\pi$--vacuum}

From the above arguments 
we cannot, however, exclude the existence of first order 
phase transitions for other values of $\theta\neq 0$.  In    
particular, we cannot discard the existence of  CP symmetry breaking for
$\theta=\pi$. Some authors claim
the existence of a sharp first order transition at \th$=\pi$
associated to spontaneous symmetry breaking of CP 
\cite{Dashen,dv2w,Ohta,Smilga} induced by massive quarks.
In that case, the analytic continuation 
of the partition function $\CZ_\theta$ to 
complex values of \th
becomes an oscillating series and there might exist values of 
$\vartheta$ as close as possible to $\vartheta=0$ where the  partition 
function $\CZ_\theta=0$ vanishes.
However, in  pure Yang-Mills theory or theories with different
quark masses some arguments support a CP symmetry preserving scenario
for the \th$=\pi$ vacuum. 
They are based on the existence of a strong  level repulsion 
between parity even and parity odd lowest energy states \cite{aff1}
The existence of such a mechanism can be derived form the analysis
of the nodal structure of the vacuum functional of $\theta=\pi$ theory. 
The study of nodal structure of vacuum functionals in quantum gauge 
theories was initiated by Feynman \cite{Feynman}. He argued that
confinement in $2+1$ dimensional Yang-Mills theories 
is a consequence of the absence of nodes and some extra properties
of the vacuum.

Standard minimum principle arguments disfavor the appearance of nodes
in  vacuum states of Quantum Theories \cite{Feynman}.
However, the presence of CP violating interactions
invalidates the use of such arguments  and the vacuum
response to this kind of  interactions can involve the
appearance of nodes. In some cases, the infrared behavior of
the theory is so dramatically modified by the CP violating
interaction that a confining vacuum state can become non-confining.
The connection between the absence of confinement
and the existence of nodes in the vacuum state, suggests that
new classical field configurations related to the nodal
structure of the quantum vacuum emerge as  new candidates to play a 
significant role in the mechanism of confinement. This idea has been
successfully exploited to show the absence of spontaneous breaking
of CP symmetry at $\theta=\pi$ for various field theories
\cite{aff1} \cite{cpn}.

In the canonical formalism and Schr\"odinger representation
the physical states of the Yang-Mills theory for any value of \th   
are wave functionals $\psi([A])$ defined on the gauge orbit space $\CM$. 
This is a consequence of Gauss law. There is no topological 
reason to force  physical states to  present 
nodal configurations, i.e. gauge orbits where the corresponding
wave functionals vanish. However, there is a very simple dynamical argument
showing that for any stationary state, including the \tth--vacuum,
there exists a physical state with nodal configurations.

The non-trivial effect of the $\theta$--term is due to the
non-simply connected character of the orbit space $\pi_1(\CM)=\IZ$
or what is equivalent the non-connected character of the group of
gauge transformations $\pi_0(\CG)=\IZ$.
It is possible, however, to perform a multivalued transformation of 
physical states
\be
{\xi(A)={\rm e}^{-{i\theta\over 2\pi}\cs}  \psi_{\rm ph}([A])}
\label{simil}
\ee
with
$$C_s(A)= 
{1\over 4\pi}\int \tr\ \left(A\wedge dA+{2\over 3}A\wedge A\wedge 
A\right),$$ 
which removes the \th dependence of the Hamiltonian
\vskip-.2cm
\be
 \IH_\theta = - {g^2\over 2} \left|\left|{\delta\over
\delta A} - {i\theta\over 8
\pi^2}\ast F(A)\right|\right|^2
 + {1\over 2 g^2} || F(A)||^2
\nonumber
\ee
$$
{\widetilde{\IH}_\theta^{\rm }=
{\rm e}^{-{i\theta\over 2\pi}\cs}
\IH_\theta^{\rm }\ {\rm e}^{{i\theta\over 2\pi}\cs}=\IH_0^{\rm
}}.$$
The transformation becomes single valued when restricted  to
the open  dense  subset 
$
{\CN=\{[A]\in\CM; C_{\rm S}(A)\neq (2n+1)\pi\},}
$
of the orbit space $\CM$.
The $\theta$
dependence is encoded in the non-trivial boundary conditions that
the new wave functionals $\xi$ satisfy at the boundary of $\CN$:
\be
{\xi(A_+)={\rm e}^{-{i\nu(\phi)\theta}} \xi(A_-)}
\label{bc}
\ee
for any pair of gauge fields
$A_-$ and $A_+$ in the same orbit of $\CM\,\backslash\, \CN$
 which are gauge equivalent
 $ A_+=A_-^\phi$  by means of a gauge transformation with 
winding number $\nu(\phi)$.
In this sense the transformation (\ref{simil})  is trading the
$\theta$--dependence of the Hamiltonian by non-trivial boundary
conditions
on $\CM\,\backslash\, \CN$. 

There are physically interesting configurations 
in this boundary domain, e.g. sphalerons.
{Sphalerons} are static solutions of Yang-Mills equations which 
by Derrick theorem can
only exist for finite space volumes. They are unstable
and become characterized by the existence of a finite number of
unstable decaying modes. The value of the Yang-Mills functional on
them marks the height of the potential barrier between classical
vacua and, therefore,  it is related to the transition temperature
necessary for the appearance of direct coalescence between those
vacua. Their existence was  predicted by  
Manton \cite{man} by an argument based
on the  non-simply connected nature of the orbit space of 
3-dimensional gauge  fields and a generalization of the
Ljusternik-Snierelman theory.  An explicit expression for the
sphaleron on $S^3$ can be obtained from the observation that the
pullback of one instanton to a 3-dimensional sphere embedded on
$S^4$ with origin  on the center of the instanton and the same
radius that the instanton is an unstable critical point of the
3-dimensional Yang-Mills functional. For $SU(2)$ the sphaleron in
stereographic coordinates reads \cite{pvB}
\vskip-.4cm 
\be 
A_j= {4\rho(4\rho \epsilon^a
_{jk} x^k - 2x^a x_j+ [x^2-4\rho^2]\delta^a_j)\sigma_a
\over (x^2+4 \rho^2)^2}.
\nonumber
\ee
\vskip-.1cm \noindent
 The unstable mode can be identified with the variation of 
the configuration under scale transformations. 

Relevant properties of sphalerons are that they give a very
special value to the Chern-Simons functional 
$C_s(A_{\hbox{\sevenpt{sph}}})=\pi$ and that they are quasi-invariant 
under parity
transformations
\vskip-.1cm
$$A^{\hbox{\sevenpt{P}}}_{\hbox{\sevenpt{sph}}}=
A^\varphi_{\hbox{\sevenpt{sph}}},$$
\vskip-.1cm\noindent
where $\varphi$ denotes the gauge transformation 
$${\varphi(x)= {1\over x^2 + 4\rho^2}\left[  (x^2 - 4\rho^2)I - 
  {4 i\rho} x^j \sigma_j\right].}
$$
with winding number
$\nu(\varphi)=1$.
In contrast, the  vacuum configuration 
$A_{{\hbox{\sevenpt{vac}}}}=0$ gives rise to a vanishing
value for the Chern-Simons functional 
$C_s(A_{{\hbox{\sevenpt{vac}}}})=0$ and is strictly
invariant under parity
$$A^{\hbox{\sevenpt{P}}}_{{\hbox{\sevenpt{vac}}}}=A_{{\hbox
{\sevenpt{vac}}}}.$$

In the case $\theta=\pi$ the boundary condition (\ref{bc})  becomes
an anti-periodic boundary condition ${\xi(A_+)=- \xi(A_-)}$  
for any pair of gauge fields
$A_-$ and $A_+$  with $[A_-]=[A_+]\in\CM\,\backslash\, \CN$
 which are gauge related  by a gauge transformation with odd
winding number.
Due to the special boundary condition  it is 
easy to see that CP parity
even states must vanish on the orbits of  sphaleron configurations 
$[A_{\hbox{\sevenpt{sph}}}]$ whereas that CP parity
odd states vanish on the orbits of classical vacuum configurations 
$[A_{{\hbox{\sevenpt{vac}}}}]$.
If the  $\pi$--vacuum state were degenerate there will exist two 
different vacuum states  $\psi^+_0$ and $\psi^-_0$ with  even  and  odd
parities, respectively.  But, since the potential term
$$V(A)={1\over 2g^2}||F(A)||^2 \phantom{\Bigg [}$$ 
of the Hamiltonian
vanishes for the vacuum configuration whereas it gets a non-null positive
value for the sphaleron gauge field, the lowest energy eigenstates
$\psi_0^+$  and $\psi_0^-$ cannot have the same energy. In this
way the  the potential term induces a level repulsion which increases
with space volume.
The splitting of those levels means that for $\theta=\pi$ there is not 
spontaneous breaking of parity, the vacuum is even and vanishes for 
sphaleron gauge fields.

Since the theory is expected to deconfine
for $\theta=\pi$, the result suggest that those
nodes might be responsible for the confining properties of the vacuum in
absence of $\theta$ term where the vacuum has no classical nodal 
configurations.

As in the case $\theta=0$ \cite{agas} the above result 
shows that CP symmetry is not spontaneously broken  also
for $\theta=\pi$ in absence of quarks
although the physical reasons are fairly different.

\section{Conclusions}

The above results  can also be extended to
other theories with similar  properties. 
One particularly interesting case is the \cp\ non-linear    
sigma model. Indeed, from a similar analysis one can conclude that    
there is no spontaneous CP symmetry breaking in \cp\ models at    
$\theta=0$ \cite{agas} and  $\theta=\pi$ \cite{cpn}.  
This result can be checked for the \cpa\ model where the    
exact solution is known for $\theta=0$ \cite{zamm,pw}. An exact    
solution which is parity preserving is also known for $\theta=\pi$ 
\cite{zammm}. In both cases the exact results are in agreement
with the results discussed above.
    
Another case where an exact solution is also known is the \cp\ model
in the large $N$ limit.  In this case the system 
describes a weakly interacting and parity preserving massive scalar
particle in the adjoint representation of $SU(N)$ at $\theta=0$ and
in the fundamental representation of $SU(N)$ at $\theta=\pi$.  The later
case has been controversial but a recent analysis \cite{mag} shows 
that it exhibits  a behavior similar to  that of Yang-Mills theory.

In conclusion, the vacuum energy density is smooth at first order for
CP symmetric vacua: $0$--vacuum and $\pi$--vacuum. The smooth behavior
qualitatively favors  cosmological applications but  quantitative 
numerical estimates still require more stringent bounds on $\theta$ 
than those based on the current data of particle physics.

\vskip 3mm
I thank M. Aguado and F. Falceto for collaboration and
J. Barb\'on, V. Laliena and E. Seiler for discussions. This
work has been partially supported by the Spanish MCyT grant
FPA2000-1252.


\begin{thebibliography}{999}   
    
\bibitem{edn}{M.~Pospelov and A.~Ritz,
Nucl.~Phys.~{\bf 573} (2000) 177};
 C.~T.~Chan, E. M.~Henley, T.~Meissner,
hep-ph/9905317

\bibitem{perl} S.~J.~Perlmutter {\it et~al}, 
Astrophys.~J. {\bf 517} (1999) 565; 
A.~G.~Riess {\it et~al}, Astron.~J. {\bf 116} (1998) 1009

\bibitem{wein} S.~Weinberg, Mod.~Phys.~Rev.,{\bf 61} (1988) 1

\bibitem{yokoyama}{J.~Yokoyama,
Phys.~Rev.~Lett. {\bf 88} (2002) 151302 ;
Int.~J.~Mod.~Phys. {\bf D 11} (2002) 1603 }

\bibitem{muz}{P.~Jaikumar, and A.~Mazumdar, 
Phys.~Rev.~Lett. (2003)}

\bibitem{vw}{C.~Vafa and E.~Witten,
Phys.~Rev.~Lett.~{\bf 53} (1984) 535}
    
\bibitem{ag}    
{V.~Azcoiti and A.~Galante, 
Phys.~Rev.~Lett.~{\bf 83} (1999) 1518 }
    
\bibitem{ch} {T.~D.~Cohen,
 Phys.~Rev.~{\bf D 64}
(2001) 047704}

\bibitem{jw} M.~B.~Einhorn, J.~Wudka, 
{\tt hep-ph/0205346} 

\bibitem{Ji} X.~Ji,
Phys.~Lett.~{\bf B 554}(2003) 33 
    
\bibitem{amm}{M.~Asorey and P.~K.~Mitter,
Commun.~Math.~Phys.~{\bf 80} (1981)
43 }

\bibitem{aff} {M.~Asorey and F.~Falceto, 
 Nucl.~Phys.~{\bf B 327} (1989) 427}

\bibitem{es}{E.~Seiler, 
{\it Gauge theories as a problem of constructive quantum
field theory and statistical mechanics}, Lect.~Notes~Phys.~{\bf 159}
(1982) 1; K.~Osterwalder, E.~Seiler,
Ann.~Phys.~ (N.Y.) {\bf 110} (1978) 440}
    
\bibitem{am} {M.~Asorey and P.~K.~Mitter,
CERN TH-2423 preprint (1982)}

\bibitem{agas}M.~Aguado and M.~Asorey,
{\tt hep-th/0204130}

\bibitem{ly}{C.~N.~Yang and T.~D.~Lee, 
Phys.~Rev.~{\bf 87} (1952) 404; 
T.~D.~Lee and C.~N.~Yang, 
Phys.~Rev.~{\bf 87} (1952) 410}
    
\bibitem{dv}  L.~Del~Debbio, H~Panagopoulos, E.~Vicari
JHEP 0208 (2002) 44

\bibitem{Dashen}
R.~F.~Dashen,  Phys.~Rev. {\bf D 3} (1971) 1879

\bibitem{dv2w} P.~Di~Vecchia and G.~Veneziano, Nucl. Phys. 
{\bf B171}(1980) 253; E.~Witten, Ann. Phys. {\bf 128} (1980) 
363

\bibitem{Ohta}N.  Ohta, Prog. Theor. Phys. {\bf 66} 
(1981) 1408; {\bf 67} (1982) 993;
M.~Creutz, Phys.~Rev. {\bf D52} (1995) 2951;
N. Evans, S. D. Hsu, A. Nyffeler and M. Schwetz,
Nucl.~Phys. {\bf B494 }(1997) 200

\bibitem{Smilga} A.~Smilga,
Phys.~Rev. {\bf D 59} (1999) 114021;
M.~H.~G.~Tytgat, Phys.~ Rev. {\bf D61}(2000) 114009;
G. Akemann, J. T. Lenaghan and K.~Splittorff,
Phys. Rev. {\bf D 65} (2002) 085015

\bibitem{aff1} {M.~Asorey and F.~Falceto, 
Phys.~Rev.~Lett.~{\bf 77} (1996) 3074}

\bibitem{Feynman} R.~Feynman, Nucl.~Phys. {\bf B188 }
(1981) 479

\bibitem{cpn} M.~Asorey and F.~Falceto, Phys.~Rev.~Lett. 
 {\bf 80} (1998) 234
  
\bibitem{man}{N.~Manton, Phys.~Rev. {\bf D28}(1983) 2019;
F.~R.~Klinkhamer and N.~Manton, Phys.~Rev. {\bf D30}(1984) 2212}

\bibitem{pvB}
{J.~M.~Cerver\'o and L.~Jacobs, { Phys.~Lett.}
{\bf B 78} (1978) 427;  
P.~van~Baal and  N. D. Hari~Dass, { Nucl.~Phys.}
{\bf B 385} (1992) 185,;
A.~V.~Smilga, { Nucl.~Phys.} {\bf B 459} (1996) 263}
    
\bibitem{zamm} {A.~Zamolodchikov and A.~B.~Zamolodchikov, 
Ann.~Phys.~(N.Y.) {\bf 120} (1979) 253}
    
\bibitem{pw} {A.~Polyakov and P.~B.~Wiegmann, 
Phys.~Lett. {\bf B131} (1983) 121}
    
\bibitem{zammm} {A.~Zamolodchikov and A.~B.~Zamolodchikov, 
Nucl.~Phys. {\bf B379}(1992) 602}

\bibitem{mag} {M.~Aguado and  M.~Asorey, In preparation}    

\end{thebibliography}
\end{document}